\documentclass[journal]{IEEEtran}

\usepackage{cite}
\usepackage{graphicx}
\usepackage{psfrag}
\usepackage{subfig}
\usepackage{stfloats}
\usepackage{amsmath}
\usepackage{array}
\usepackage{topcapt}
\usepackage[nolineno]{lgrind}
\usepackage{algorithm}
\usepackage{algorithmic}

\usepackage{color}

\centerfigcaptionstrue 

\begin{document}

\title{Rapid Prototyping over IEEE 802.11}

\author{Fehmi Ben Abdesslem, Luigi Iannone, Marcelo Dias de Amorim, Katia
Obraczka, Ignacio Solis, and Serge Fdida
\thanks{Fehmi Ben Abdesslem, Marcelo Dias de Amorim, and
Serge Fdida are with the Laboratoire d'Informatique de Paris 6
(LIP6/CNRS), Universit{\'{e}} Pierre et Marie Curie~-- Paris 6,
France. Email: \{fehmi, amorim, sf\}@rp.lip6.fr. Luigi Iannone is
with the D{\'{e}}partement d'Ing{\'{e}}énierie Informatique,
Universit{\'{e}} Catholique de Louvain, Belgium. Email:
iannone@info.ucl.ac.be. Katia Obraczka is with the  Computer
Engineering Department, University of California, Santa Cruz, USA.
Email: katia@cse.ucsc.edu. Ignacio Solis is with Palo Alto Research
Center, USA. Email: isolis@parc.com. This work has been partially
supported by the IST European project WIP under contract 27402 and
by the RNRT project Airnet under contract 01205.}}

\thispagestyle{plain}


\maketitle

\begin{abstract}
This paper introduces Prawn, a tool for {\em rapid prototyping}
communication protocols over IEEE~802.11 networks. Prawn provides a
software environment that makes prototyping as quick, easy, and
effortless as possible and thus allows researchers to conduct both
functional assessment and performance evaluation as an integral part
of the protocol design process. Since Prawn runs on real IEEE~802.11
nodes, prototypes can be evaluated and adjusted under realistic
conditions. Once the prototype has been extensively tested and
thoroughly validated, and its functional design tuned accordingly, it
is then ready for implementation. Prawn facilitates prototype
development by providing: (i) a set of building blocks that implement
common functions needed by a wide range of wireless protocols (e.g.,
neighbor discovery, link quality assessment, message transmission and
reception), and (ii) an API that allows protocol designers to access
Prawn primitives. We show through a number of case studies how Prawn
supports prototyping as part of protocol design and, as a result of
enabling deployment and testing under real-world scenarios, how Prawn
provides useful feedback on protocol operation and performance.
\end{abstract}

\section{Introduction}
\label{sec:introduction}

Designing protocols for wireless networks poses countless technical
challenges due to a variety of factors such as node mobility, node
heterogeneity, power limitations, and the fact that the
characteristics of the wireless channel are non-deterministic and
can be highly variant in space and time. This implies that testing
and evaluating such protocols under real operating conditions is
crucial to ensure adequate functionality and performance.

In fact, the networking research community has already acknowledged
the importance of testing and evaluating wireless protocol proposals
under real-world conditions. As a result, over the last few years, a
number of testbeds, such as Orbit~\cite{Bib:Raychaudhuri05},
UnWiReD's testbed~\cite{Bib:Zhu05}, Netbed~\cite{Bib:White03}, and
Roofnet~\cite{Bib:De06,Bib:roofnet}, as well as implementation
tools, such as Click~\cite{Bib:Kohler00} and
XORP~\cite{Bib:Handley03}, have been developed to support the
deployment and evaluation of wireless protocols under realistic
scenarios.

\begin{figure}[t]
 \begin{center}
 \subfloat[Traditional design cycle.]{
     \label{fig:no-proto}
     \includegraphics[width=\linewidth]{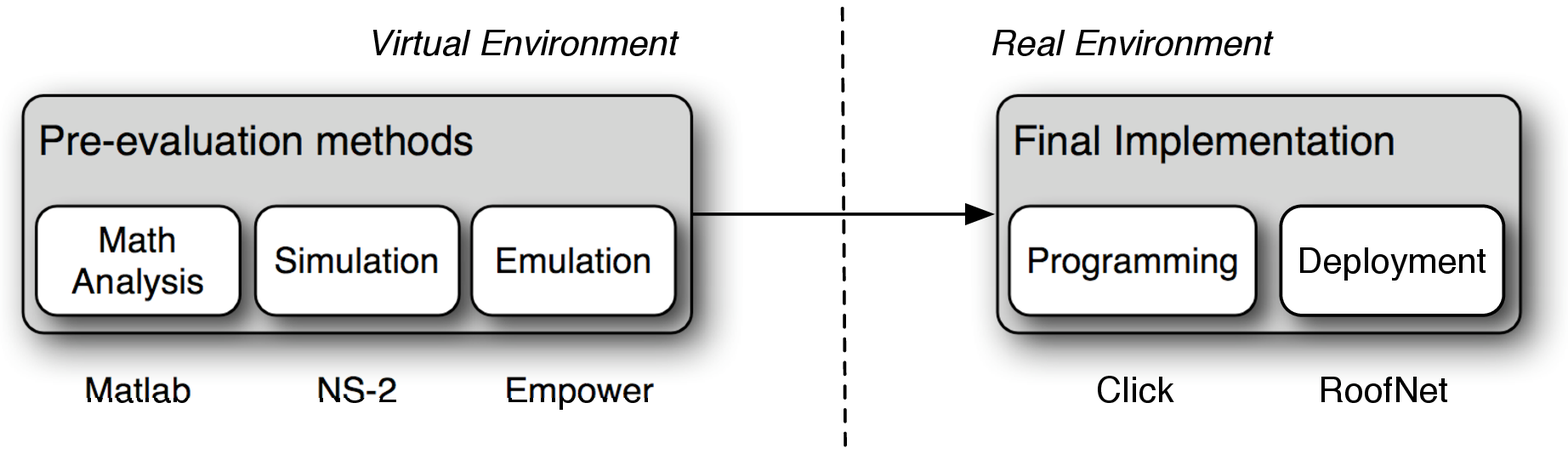}
 }\\
\subfloat[Design cycle including rapid prototyping.]{
    \label{fig:proto}
    \includegraphics[width=\linewidth]{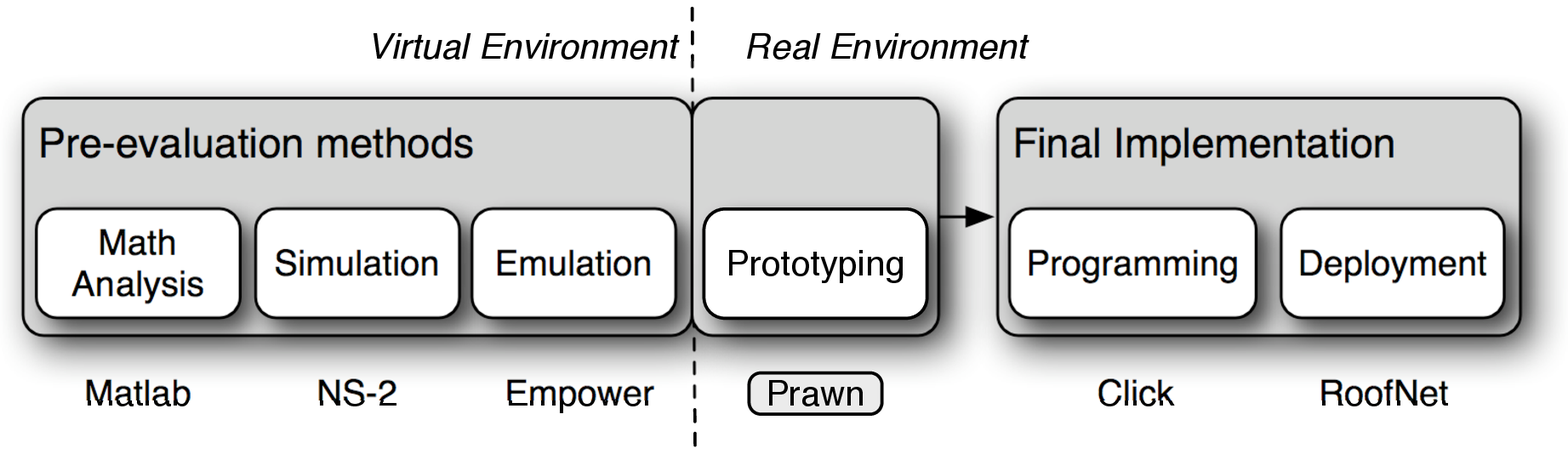}
} \caption{Bridging the gap between theory and practice in the
design of protocols and systems for wireless networks.}
\label{fig:feedback-loop}
\end{center}
\end{figure}

As illustrated in Figure~\ref{fig:no-proto}, there are mainly three
evaluation methodologies commonly used when designing communication
systems, namely mathematical analysis, simulation, and emulation. In
this paper, we go a step further and advocate including {\em rapid
prototyping} as an integral part of the design process (cf.,
Figure~\ref{fig:proto}). This will enable performing correctness
verification, functionality and performance tests under real operating
conditions early enough in the design cycle that resulting feedback
and insight can be effectively incorporated into the design. Rapid
prototyping is complimentary to current testbeds and tools which are
typically used to produce a beta version of the final implementation,
a step just before public release. Therefore, testing a protocol under
real conditions often happens at the end of the development cycle or
even after it is over.

We postulate that what is needed is a tool that makes prototyping as
quick, easy, and effortless as possible. To this end, we introduce {\it
Prawn} (PRototyping Architecture for Wireless Networks), a novel
software environment for prototyping high-level (i.e., network layer
and above) wireless network protocols. Prawn's approach to rapid
prototyping is based on two main components:

\begin{itemize}

\item The {\bf Prawn Engine}, a set of basic building blocks atop
which protocols and services can be prototyped. These building
blocks include functions such as neighbor discovery, link
assessment, and device configuration.

\item The {\bf Prawn Library}, an API that provides protocol
designers with easy and transparent access to the underlying
building blocks. Prawn deliberately provides a concise set of
communication primitives, yet sufficient for a wide range of
high-level wireless protocols.

\end{itemize}

Prototypes implemented with Prawn are not expected to be optimized,
offering edge performance. Rather, our focus with Prawn is on
obtaining, quickly and with little effort, a complete and fully
functional instantiation of the system. Prawn makes prototyping as
simple as writing network simulation scripts, with the difference
that testing is done under realistic conditions.\footnote{Currently,
Prawn targets IEEE 802.11 networks, although its design can be
extended to run atop other wireless network technologies.} Assessing
these conditions is done through the Prawn Engine, which runs as a
background process that proactively performs tasks such as neighbor
discovery and link quality assessment. This feature allows Prawn to
provide accurate and up-to-date feedback from the wireless
interface.

As shown by the several case studies presented in this paper, Prawn
prototypes can be used for functional assessment as well as both
absolute and comparative performance evaluation.  Once the prototype
has been extensively tested and thoroughly validated, and its
functional design tuned accordingly, it is then ready for final
implementation (which is out of the scope of Prawn).

In summary, Prawn's contributions are as follows.

\begin{enumerate}

   \item {\it Prawn enables rapid prototyping} which is key to testing
   and evaluating network protocols and services under real operating
   conditions as early as possible in the design cycle.

   \item {\it Prawn provides an easy-to-use and extensible prototyping
   interface} which makes prototyping considerably simpler and faster
   requiring only basic programming skills. Prawn is to ``live''
   experimentation what scripts are to simulations. Furthermore, new
   primitives can be easily incorporated as needed.

   \item {\it Prawn includes a flexible active neighborhood
   discovery mechanism} which provides configurable neighborhood
   probing and power control mechanisms.

\end{enumerate}

The remainder of this paper is organized as follows. We put our work
on Prawn in perspective by reviewing related work in the next section.
Section~\ref{sec:overview} provides an overview of Prawn, while in
Sections~\ref{sec:library} and~\ref{sec:engine} we describe Prawn's
two main components in detail. We evaluate the overhead introduced by
Prawn in Section~\ref{sec:overhead} and present in
Section~\ref{sec:case} a number of case studies showing how Prawn
makes prototyping fast and simple. Finally, we present our concluding
remarks and directions for future work in
Section~\ref{sec:conclusion}.

\section{Related Work}
\label{sec:related}

Simulations are perhaps the most widely used methodology for
evaluating network protocols. They allow designers to evaluate the
system at hand under a wide range of conditions (e.g., different
mobility models, node heterogeneity, varying channel conditions). They
also allow the exploration of the design space by enabling designers
to vary individual protocol parameters (e.g., timers) and combinations
thereof. Finally, they are instrumental for scalability analysis and
they offer reproducibility. Examples of well known simulation
platforms include NS-2~\cite{Bib:NS-2}, OPNET~\cite{Bib:OPNET},
GloMoSim~\cite{Bib:Glomosim}, and QualNet~\cite{Bib:Qualnet}.

Emulation tries to subject the system under consideration to real
inputs and/or outputs. Environments like EMPOWER~\cite{Bib:Zheng03}
or Seawind~\cite{Bib:Kojo01} emulate the wireless medium by
introducing packet error rates and delays. Other emulators like
m-ORBIT~\cite{Bib:Ramachandran05} also emulate node mobility by
space switching over a testbed of fixed nodes. A key advantage of
emulation in the context of wireless/mobile networks is to
facilitate testing by avoiding, for example, geographic and mobility
constraints required for deployment.


More recently, a number of projects have pioneered the field of
wireless protocol evaluation under real conditions.\footnote{Given
the focus of this paper, we highlight related efforts that target
wireless networks. However, similar tools for Internet research have
been proposed; notable examples include
VINI~\cite{Bib:bavier.sigcomm06}, PlanetLab~\cite{Bib:PlanetLab},
X-Bone~\cite{Bib:touch.global98}, and
Violin~\cite{Bib:jiang.ispdpa04}.} They include testbeds such as
Orbit~\cite{Bib:Raychaudhuri05}, Emulab~\cite{Bib:eide.nsdi07},
Roofnet~\cite{Bib:roofnet}, Mint-m~\cite{Bib:De06}, the work
reported in UnWiReD~\cite{Bib:Zhu05} and Netbed~\cite{Bib:White03},
as well as tools that support protocol implementation like the Click
modular router~\cite{Bib:Kohler00} and XORP~\cite{Bib:Handley03}).
As previously pointed out, such tools and Prawn have different
goals, address different phases of the design process, and are
therefore complementary. While tools like Click and XORP targets the
final implementation at the final stages of protocol design, Prawn
focuses on prototyping a research proposal at the very early stages
of the design process. Therefore, through Prawn, protocol designers
can very quickly and easily generate a fully functional, but
non-optimized, implementation for live testing in real scenarios.

In the context of wireless sensor networks (WSNs), Polastre \emph{et
al.}~\cite{Bib:Polastre05} propose SP (Sensornet Protocol), a
unifying link abstraction layer. SP runs on TinyOS~\cite{Bib:Tinyos}
and provides an interface to a wide range of data-link and physical
layer technologies. Prawn and SP roughly share the same functional
principles, e.g., data transmission, data reception, neighbor
management with link quality, etc. However, they have quite
different goals. First, SP only manages the neighbor table; it
neither performs neighbor discovery nor provides link assessment.
Second, SP is designed for WSNs whereas Prawn is for general
IEEE~802.11 networks. Finally, while SP aims at optimizing the
communication and unifying different link layers in WSNs, Prawn aims
at facilitating and simplifying prototype implementation.

EmStar~\cite{Bib:Girod04} is another development environment for
WSNs and runs on the Intel Stargate platform~\cite{Bib:XBow2}. It is
similar in essence to Prawn since it provides a set of primitives
that upper layers can use. However, EmStar supports implementation
by focusing on modularity and code reuse. Its architecture is quite
complex and its use requires quite sophisticated development skills
when compared to Prawn standards.

MAPI~\cite{Bib:Mapi} is an API especially developed for wireless
networks based on the Wireless Tools~\cite{Bib:Jt}. It provides a
set of simple primitives to obtain information from the underlying
wireless device. Besides accessing information from the wireless
device through a simple API, Prawn also runs an active daemon that
performs neighborhood discovery with link quality assessment, as
well as sending/receiving mechanism.

%

\section{Prawn Overview}
\label{sec:overview}

Prawn targets prototyping protocols and services at the network
layer and above.  Simplicity was a major goal we had in mind when
designing Prawn; we wanted to ensure that learning how to use Prawn
would be as intuitive and immediate as possible requiring only basic
programming expertise. For example, in Prawn, sending a packet at a
given transmit power level is performed by a single primitive and
takes one line of code. Our focus was thus to provide: (1) a concise,
yet complete set of functions to realize high-level protocols and (2)
a simple, easy-to-use interface to provide access to Prawn's
functionalities.

\subsection{Prawn Architecture}
\label{subsec:architecture}

Prawn consists of two main components: (i) the
Prawn Library (cf., Section~\ref{sec:library}), which provides
high-level primitives to send and receive
messages, retrieve information from the network, etc; and (ii) the
Prawn Engine (cf., Section~\ref{sec:engine}), which implements the
primitives provided by the Prawn Library.

The current implementation of Prawn runs on Linux atop IP
for backward compatibility with the global Internet. The interaction
between the Prawn Engine and the physical wireless device relies on
the Wireless Tools~\cite{Bib:Jt}. This set of tools allows retrieving
information from most wireless devices as well as setting low-level
parameters. Furthermore, it is available with most Linux distributions.

\begin{figure}[t!]
\centering
\includegraphics[width=7cm]{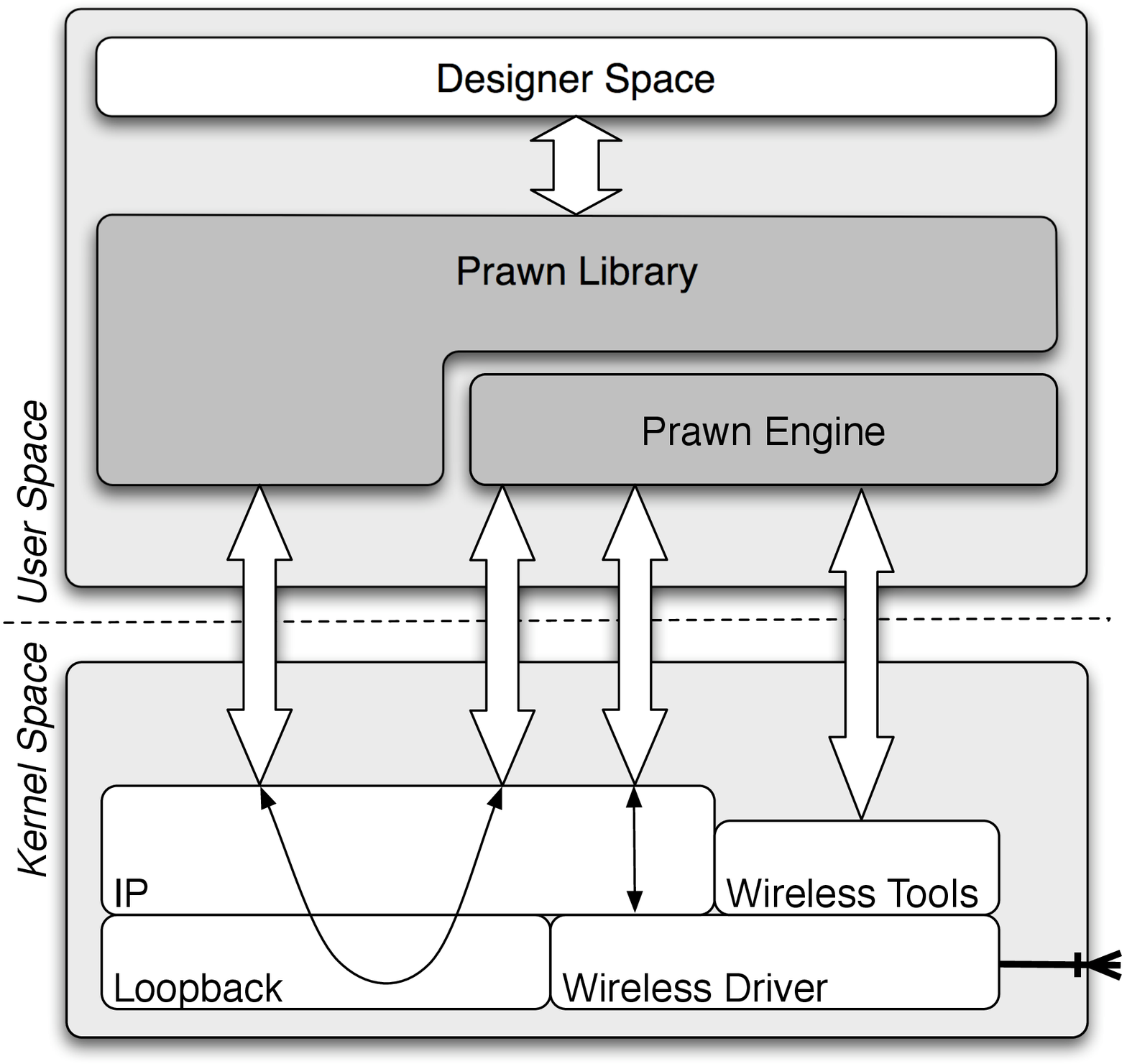}
\caption{Prawn architecture} \label{fig:architecture}
\end{figure}

Prawn's components, how they interact with one another and with the
underlying operating system are illustrated in
Figure~\ref{fig:architecture}. As highlighted in the figure,
Prawn's functionalities are accessible through the Prawn
Library. Messages and requests received from the library are then
processed by the Prawn Engine. The Prawn Library and Engine
communicate with each other through the loop-back interface using a
simple request/reply mechanism. This choice simplifies modularity
and portability.

\subsection{How to Use Prawn}
\label{subsec:howto}

Running Prawn requires only a few basic steps. First, it needs to be
configured and installed on the machines that will be used in the
experiments. In particular, in the Prawn configuration file it is
necessary to set the names of the wireless interface and network
(e.g., the ESSID). Optionally an IP address can be specified.
Otherwise, Prawn will randomly generate an IP address in a default
subnetwork.

Using Prawn itself only requires two operations (as described
below), namely executing the Prawn Engine and including the Prawn
Library in the prototype code.\vspace{1mm}

\noindent {\bf 1: Starting the Prawn Engine.} Prawn is distributed
under GPL license and available online~\cite{Bib:prawn}. Once
compiled, the Prawn Engine is launched as a command line program on
machines connected in ``ad hoc'' mode. Prawn is supposed to run in
daemon mode, but can run in console mode for debugging purposes. As
stated before, Prawn provides a number of options that can be
set/configured at the execution of the Engine. They are listed in
Table~\ref{tab:options}. Other options (e.g., the number of lost
beacons required to consider that a node is no more a neighbor) are
tunable in the ${\tt prawn.cfg}$ configuration file.\vspace{1mm}

\begin{table}[t!]
\topcaption{Prawn's command line options.} \label{tab:options}
\vspace{-5mm}
  \begin{center}
    \small
    \begin{tabular}{|l|c|c|}
      \hline {\bf Option} & {\bf Parameter}           & {\bf Default}\\ \hline
      \hline -N name      & node ID                   & hostname \\
      \hline -b period    & beacon period in ms       & 10000 \\
      \hline -h           & help                      & -- \\
      \hline -d           & daemon mode               & -- \\
      \hline -v           & verbose mode              & -- \\
      \hline -vv          & more verbose              & -- \\
      \hline -p port      & neighbor port             & 3010 \\
      \hline -c port      & client port               & 3020 \\
      \hline -i I         & uses wireless interface I & ath0 \\
      \hline -P           & set transmit power level  & -- \\
      \hline -n           & no power control features & -- \\
      \hline -W window    & window size for PER       & 5 \\
      \hline -V           & version                   & -- \\
      \hline
    \end{tabular}
  \end{center}
\end{table}

\noindent {\bf 2: Using the Prawn Library.} The Prawn Library
(described in detail in Section~\ref{sec:library}) is composed of a
set of primitives that are linked to the prototype through standard
include files. Currently, prototypes can be developed either in C or
in Perl (a Java version is about to be released). For C development,
the file ${\tt prawn.h}$ should be included in the header of the
prototype code. Similarly, the file ${\tt prawn.pl}$ is to be included
for prototypes developed in Perl.

\subsection{``Hello World!''}
\label{subsec:hello}

To illustrate the use of Prawn, we describe how to implement a simple
``hello world'' prototype using Prawn's Perl library. In this
example we send a message from Bob to Alice.\vspace{1mm}

\noindent {\bf Step 1.} Launch Prawn with ``{\tt prawn -d -N Bob}''
in the first machine and ``{\tt prawn -d -N Alice}'' in the second
machine.\vspace{1mm}

\noindent {\bf Step 2.} Get the first machine ready to receive
messages by executing the following Perl script:
\vspace{2mm}

\begin{minipage}[h]{0.9\linewidth}

\Head{}
\File{receive.pl}{2007}{1}{28}{16:34}{129}
\L{\LB{\V{require}_\S{}\3prawn.pl\3\SE{};}}
\L{\LB{\K{while}(!@\V{Message})\{}}
\L{\LB{@\V{Message}=\V{Prawn\_Receive}();}}
\L{\LB{\}}}
\L{\LB{\K{print}_\S{}{'}Received_:_{'}\SE{}.\N{\$Message}[\N{4}].\S{}{'}_from_{'}\SE{}.\N{\$Message}[\N{2}].\S{}\3\2n\3\SE{};_}}

\end{minipage}

\noindent {\bf Step 3.} On the other machine launch the
following Perl script:
\vspace{2mm}

\begin{minipage}[h]{0.9\linewidth}

\Head{}
\File{send.pl}{2007}{1}{28}{16:34}{50}
\L{\LB{\V{require}_\S{}\3prawn.pl\3\SE{};}}
\L{\LB{\V{Prawn\_Send}(\S{}\3Hello World!\3\SE{},\S{}\3Bob\3\SE{});_}}

\end{minipage}

The result is trivial: ${\tt Alice}$ sends a ``Hello World'' message
to ${\tt Bob}$, and ${\tt Bob}$ prints ``{\tt Received: Hello World
from Alice}'' on the screen. However, this simple example aims at showing the
level of abstraction provided by Prawn, where low-level system
knowledge (e.g., sockets, addressing) is required. More elaborated
examples will be presented in Section~\ref{sec:case}.

\section{The Prawn Library}
\label{sec:library}

The Prawn Library, currently implemented in C and Perl, provides a set
of high-level communication-oriented functions. They hide from
protocol designers lower-level features such as addressing,
communication set-up, etc. Their syntax is quite simple and intuitive.
Prawn's current set of primitives addresses basic functions required
when prototyping a high-level communication protocol; nevertheless,
Prawn was designed to be easily extensible allowing new primitives to
be implemented and integrated. The primitives currently available
are:\vspace{1mm}

\begin{itemize}

\item ${\tt Prawn\_Info()}$: Returns information on the
configuration of the local Prawn Engine. Basically, it consists of
the list of settings chosen when launching the daemon (cf.,
Table~\ref{tab:options}). Some examples are the node's ID, interface
port number, and beacon period.

\item ${\tt Prawn\_Neighbors()}$: Returns the list of the node's
one-hop and two-hop neighbors as well as statistics concerning the
quality of the respective links. In Section~\ref{sec:engine}, a
thorough explanation of the information returned by the Engine will
be given.

\item ${\tt Prawn\_Send(Message,\, ID,\, TX\_Pwr)}$: Sends ${\tt Message}$
to node ${\tt ID}$; the optional argument ${\tt TX\_Pwr}$ can be
used to explicitly set the transmit power to be used during the
transmission. ${\tt Message}$ can be a string, a number, a data
structure, or any other data or control message, depending on the
prototyped protocol (e.g., a route request primitive of a route
discovery protocol).

\item ${\tt Prawn\_Send\_Broadcast(Message,\ TX\_Pwr)}$: Sends a
broadcast message containing ${\tt Message}$; in a similar way to
${\tt Prawn\_Send}()$, the optional argument ${\tt TX\_Pwr}$ allows
to set the transmit power.

\item ${\tt Prawn\_Receive()}$: Checks if a message has been
received; if so, the message is returned. This primitive is
non-blocking: if no message has been received, it just returns zero.

\end{itemize}

\section{The Prawn Engine}
\label{sec:engine}

The Prawn Engine is event-driven, i.e., its main process remains
asleep waiting for an event to occur. An event can be triggered by a
request from the Prawn Library (coming through the loop-back
interface) or by a message received on the wireless interface. The
main loop of the Engine is described in pseudo-code in
Algorithm~\ref{alg:mainloop}.  The main events are:\vspace{1mm}

\noindent {\bf Control Event:} Prawn performs some tasks on a
regular basis controlled by a timer. For instance, a timeout event
triggers the transmission of neighborhood discovery control messages
(beacons or beacon replies, cf. Sections~\ref{subsec:beacons}
and~\ref{subsec:replying}).\vspace{1mm}

\noindent {\bf Client Request:} This is an asynchronous event. It is
triggered by a library call, requesting an action from the engine
(e.g., sending a packet or retrieving the current neighbor
list).\vspace{1mm}

\noindent {\bf Neighbor Message:} This event is also asynchronous.  It
is triggered when messages from neighbors are received through the
wireless interface. These can be either control messages to be
processed by the engine or user messages to be delivered to the
prototype through the library's primitive ${\tt
Prawn\_Receive()}$.\vspace{1mm}

\begin{algorithm}[t]
\caption{The Prawn Engine's main loop.} \label{alg:mainloop}
\begin{algorithmic} [1]
\small
 \STATE Timeout$ \Leftarrow$ $time\_to\_next\_regular\_event$
 \WHILE{1}
    \IF {(Timeout)}
        \STATE Perform regular operation
        \STATE Timeout$ \Leftarrow$ $time\_to\_next\_regular\_event$
    \ELSIF{(Client Request)}
          \STATE Perform requested action
    \ELSIF{(Neighbor Message)}
          \IF{(Message == Data)}
             \STATE Send packet to the client process
          \ELSIF{(Message == Control)}
             \STATE Update neighborhood list and statistics
          \ENDIF
    \ENDIF
\ENDWHILE
\end{algorithmic}
\end{algorithm}

\subsection{Packet Format}
\label{subsec:packet-format}

All packets transmitted by Prawn start with a one-byte ${\tt Type}$
field that defines the structure of the rest of the packet. In its
current version, Prawn defines four types of packets as
shown in Table~\ref{tab:pkttypes}.

\begin{table}[h]
\topcaption{Prawn's packet types.} \label{tab:pkttypes}
\vspace{-5mm}
    \begin{center}
        {\small \begin{tabular}{|c|l|} \hline
        \textbf{Type \#}  & \textbf{Function} \\ \hline
        ${\tt 0}$         & Reserved          \\ \hline
        ${\tt 1}$         & Beacon            \\ \hline
        ${\tt 2}$         & Data              \\ \hline
        ${\tt 3}$         & Feedback          \\ \hline
        \end{tabular}}
    \end{center}
\end{table}

\subsection{Beaconing}
\label{subsec:beacons}

To build and maintain the list of neighbors, each node running Prawn
broadcasts 24-byte beacons periodically.
The beacon period is configurable depending on the requirements of the
prototype under development. By default, the Prawn Engine is
configured to test connectivity under different power levels (useful
for instance to prototype topology control algorithms based on power
control~\cite{Bib:blough.tmc06,Bib:li.ton05}). The Prawn Engine
applies a round-robin policy to continuously change the transmit
power. A beacon is first broadcast with the lowest power value. The
transmit power level is successively increased for each beacon, up to
the maximum transmit power. We call this sequence of beacons a {\it
cycle}. The different values of the transmit power are either obtained
from the interface or set by the user. This cycle is then repeated at
every beacon period. This way, the time elapsed between two beacons
sent with the same transmit power is equal to the beacon period.

Of course, the power control feature is optional, depending on the
designer's needs. If this feature is disabled, each cycle is then
composed of only one beacon, sent at the default transmit power
level. The number of transmission power levels and their values are
customizable, depending on the power control features provided by
the wireless interface under utilization.

\begin{figure}[t!]
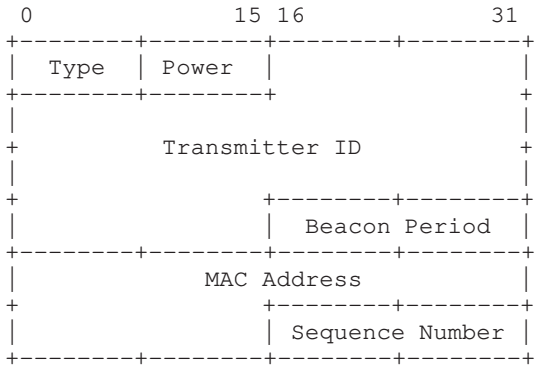

  \begin{center}
{\small \begin{verbatim}
      0              15 16             31
     +--------+--------+--------+--------+
     |  Type  | Power  |                 |
     +--------+--------+                 +
     |                                   |
     +          Transmitter ID           +
     |                                   |
     +                 +--------+--------+
     |                 |  Beacon Period  |
     +--------+--------+--------+--------+
     |             MAC Address           |
     +                 +--------+--------+
     |                 | Sequence Number |
     +--------+--------+--------+--------+
\end{verbatim}}
\caption{Prawn beaconpacket format.} \label{fig:header-beacon}
\end{center}
\end{figure}

The beaconing packet format, which is illustrated in
Figure~\ref{fig:header-beacon}, includes the following fields:
\begin{itemize}

\item ${\tt Type}$: This field is set to `{\tt 1}' (cf.,
Table~\ref{tab:pkttypes}).

\item ${\tt Transmit\ Power}$: Transmit power used to send the
beacon.

\item ${\tt Transmitter\ ID}$: Sender identifier.

\item ${\tt Beacon\ Period}$: Time period between two beacons
transmitted with the same power level (set by the user).

\item ${\tt MAC\ Address}$: MAC address of the transmitter.

\item ${\tt Sequence\ Number}$: Sequence number of the beacon.

\end{itemize}

Upon the reception of a beacon (or sequence of beacons if different
transmit powers are used), various statistics can be derived. For
instance, a node $\texttt{A}$ can determine, at a given point in time,
the minimum transmit power that $\texttt{B}$ should use to send
messages to $\texttt{A}$. This value corresponds to the lowest
transmit power among all the beacons received by $\texttt{A}$ from
$\texttt{B}$. Of course, the minimum transmit power may change over
time, and will be updated along the successive cycles.

Configuring Prawn is important to achieve an adequate balance
between performance and overhead. For example, sending beacons too
frequently would generate high overhead. On the other hand,
limiting the number of beacons is likely to result in out-of-date
measures. For these reasons, the beaconing period is one of Prawn's
customizable parameters and its value is carried in the header
of each beacon sent. A beacon is considered lost when the
beaconing period (included in previous received beacons) times out. By
default, a neighbor is removed from a node's neighbor table when
three consecutive beacons from this neighbor have been lost (or when
three consecutive beacons for every transmit power have been lost).

\subsection{Replying to Beacons}
\label{subsec:replying}

\begin{figure}[t!]
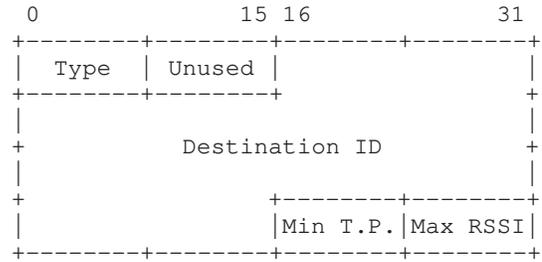

  \begin{center}
{\small \begin{verbatim}
      0              15 16             31
     +--------+--------+--------+--------+
     |  Type  | Unused |                 |
     +--------+--------+                 +
     |                                   |
     +           Destination ID          +
     |                                   |
     +                 +--------+--------+
     |                 |Min T.P.|Max RSSI|
     +--------+--------+--------+--------+
\end{verbatim}}
\caption{Prawn feedback packet format.}
\label{fig:header-feedback}
\end{center}
\end{figure}

Nodes reply to beacons using 16-byte feedback packets, as shown in
Figure~\ref{fig:header-feedback}. Feedback packets summarize
neighborhood-- and link quality information as perceived by the
receiver of the beacons. This feature allows verifying the
bidirectionality of links. Feedback packets are sent to every
neighbor after a complete cycle.\footnote{Note that if the power
control feature is disabled, then the cycle is unitary.} Prawn keeps
sending feedback packets also in the case where a neighbor is
considered lost (a unidirectional link may still exist between the
two nodes). Feedback packets contain the following fields:

\begin{itemize}

\item ${\tt Type}$: This field is set to `{\tt 3}' (cf.,
Table~\ref{tab:pkttypes}).

\item ${\tt Destination\ ID}$: Identifier of the neighbor concerned by the
feedback.

\item ${\tt Minimum\ Received\ Transmit\ Power}$: Is the transmit
power of the beacon received with the weakest signal strength from
that particular neighbor.

\item ${\tt Maximum\ Received\ Power\ Strength}$ (in dBm): Is the maximum
signal strength measured when receiving beacons from that particular
neighbor.

\end{itemize}

The rationale for reporting the transmit power of the weakest beacon
received from a particular neighbor is that it allows to roughly
characterize the quality of the corresponding link. This estimation
is also confirmed using the maximum received signal strength measured
within a cycle.

Although destined to a single neighbor, feedback packets are broadcast
and thus overheard by all one-hop neighbors. This way, nodes can
obtain information on two-hop neighborhood (cf.,
Figure~\ref{fig:list-engine}).

\subsection{Getting Information from Prawn}
\label{subsec:information}

\begin{figure}[!t]
  \begin{center}
    \includegraphics[width=\linewidth]{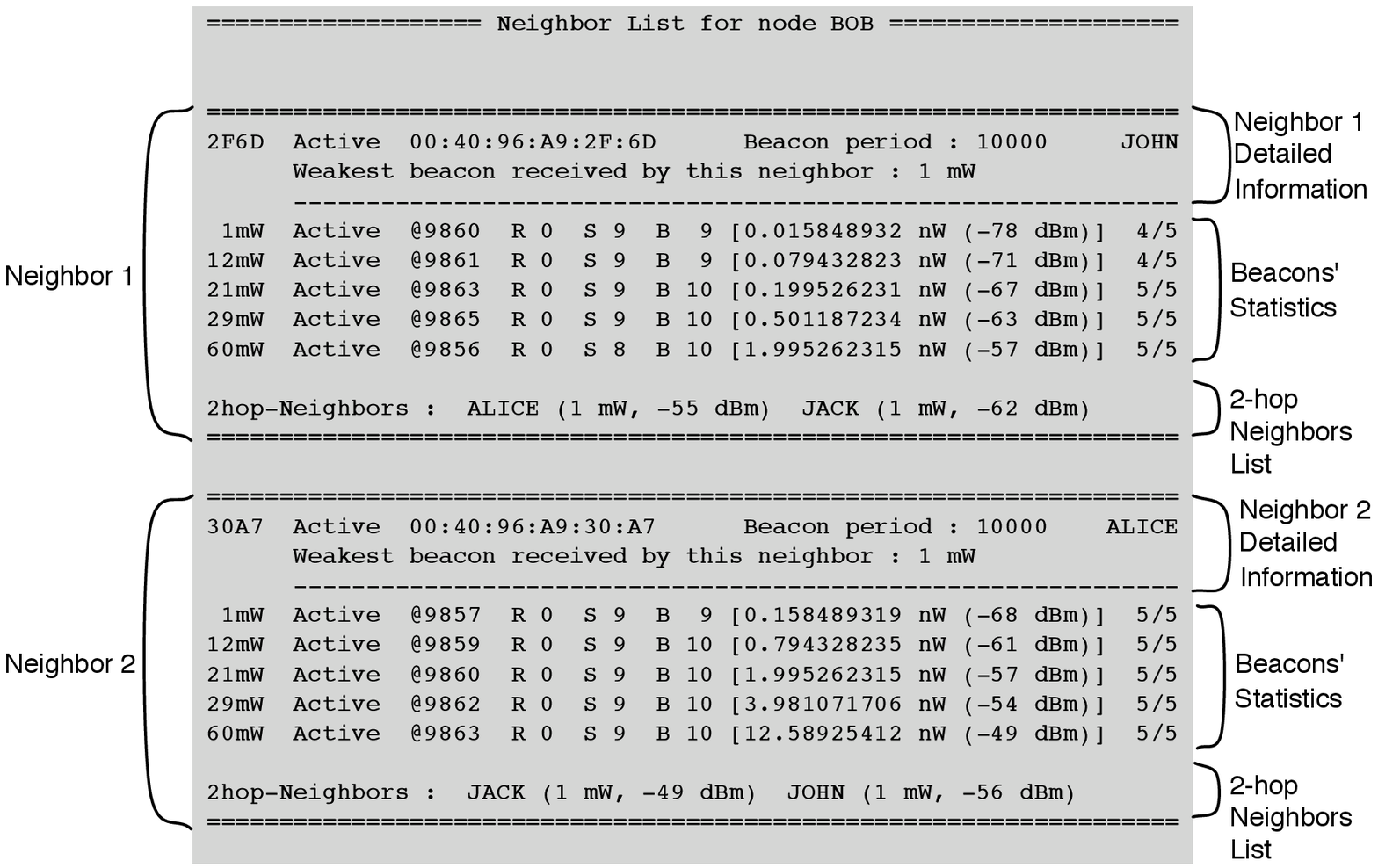}
  \caption{Information provided by Prawn for a node whose ID is ``${\tt BOB}$''.}
  \label{fig:list-engine}
\end{center}
\end{figure}

When a node calls the ${\tt Prawn\_Neighbors()}$ primitive, the
engine returns a data structure with information about the node's
neighborhood. This information can be also obtained by
running Prawn in console mode, e.g., for debugging purposes.
Figure~\ref{fig:list-engine} shows a snapshot of the information
returned by the Prawn Engine running in console mode on a node named
``${\tt Bob}$''. This snapshot shows a list of ${\tt Bob}$'s
neighbors, along with statistics on last beacons received by each
neighbor for every transmit power. Basically, ${\tt Bob}$ has two
active neighbors, ${\tt John}$ and ${\tt Alice}$. The link between
${\tt Bob}$ and ${\tt Alice}$ has, on average, better quality than
the one between ${\tt Bob}$ and ${\tt John}$; indeed for beacons
sent at 1~mW and 12~mW, only 4/5 of them have been received.

As previously described, neighborhood information is obtained
through beacons and feedback packets.  More specifically, broadcast
beacons are used to build the list of direct neighbors. This list is
established by gathering the transmitter ID of each received beacon.
Moreover, data included in beacons and feedback packets inform each
node what is the minimum transmit power required to reach a
neighbor. Such information is of primary importance in assessing
link quality. Another prominent link characteristic is the error
rate, which is determined according to the beacon period included in
each beacon transmitted. The Engine considers a beacon as lost when
it is not received within the beacon period indicated by the
corresponding neighbor. The size of the receiving window used to
compute the error rate is customizable. For instance, in
Figure~\ref{fig:list-engine}, the error rate for $\texttt{John}$'s
packets transmitted at 12~mW is $1/5$, because over the 5 most
recent 12~mW beacons transmitted by John, only 4 have been received.

When receiving a beacon, the Prawn Engine retrieves and saves the
received signal strength. Along with the transmit power of the
beacon (which is also included in the beacon), the received signal
strength returned by the engine helps to evaluate the signal
attenuation. The difference between the transmitted power level
indicated in the beacon and the signal strength measured when the
beacon is received can also be used by a protocol to characterize
link quality.

\subsection{Sending and Receiving Messages}
\label{subec:packets}

\begin{figure}[t!]
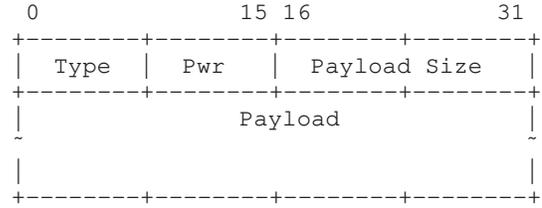

  \begin{center}
{\small \begin{verbatim}
      0              15 16             31
     +--------+--------+--------+--------+
     |  Type  |  Pwr   |  Payload Size   |
     +--------+--------+--------+--------+
     |               Payload             |
     ~                                   ~
     |                                   |
     +--------+--------+--------+--------+
\end{verbatim}}
\caption{Prawn data packet.} \label{fig:header-data}
\end{center}
\end{figure}

Two other key functions performed by the Prawn Engine are
transmission and reception of data (triggered by the ${\tt
Prawn\_Send()}$ and ${\tt Prawn\_Receive()}$ primitives,
respectively). The engine is in charge of the communication set up,
namely opening sockets, converting the receiver identifier to a
valid IP address, encapsulating/decapsulating packets, and adjusting
the transmit power before transmission.
Figure~\ref{fig:header-data} shows the structure of the data
packets, which contain the following fields.
\begin{itemize}

\item ${\tt Type}$: This field is set to `{\tt 2}' (cf.,
Table~\ref{tab:pkttypes}).

\item ${\tt Transmit\ Power}$: Power used to send the packet.

\item ${\tt Payload\ Size}$: Size of the payload field.

\item ${\tt Payload}$: Data being sent.

\end{itemize}

Data packets are sent using UDP to the corresponding IP address.
This explains why their header does not need to include the
destination ID.\footnote{Note that the same method cannot be used
for beacons, since beacons are always sent broadcast at IP level and
thus contain the broadcast address.} On the receiver side, the
engine listens on an open socket for any incoming packets. Packets
are then decapsulated and sent to the prototype which retrieves them
by using the ${\tt Prawn\_Receive()}$ primitive.

\begin{table}[t!]
{\small \topcaption{Average delays measured using Prawn on a Dell
Latitude X1 laptop.} \label{tab:laptop2} \vspace{-5mm}
   \begin{center}
       \begin{tabular}{|l|c|c|}
       \hline  & 100-byte packets & 1,400-byte packets\\
       \hline Delay to send            & 0.13~ms, $\sigma$=0.016~ms & 0.15~ms, $\sigma$=0.018~ms \\
       \hline Delay to receive         &0.14~ms, $\sigma$=0.008~ms & 0.15~ms, $\sigma$=0.008~ms\\
       \hline
       \end{tabular}
   \end{center}
   }
\end{table}

\begin{table}[t!]
{\small \topcaption{Average delays measured using Prawn on a HP
Compaq nx7000 laptop.} \label{tab:laptop1} \vspace{-5mm}
   \begin{center}
       \begin{tabular}{|l|c|c|}
       \hline \textbf{} & 100-byte packets  & 1,400-byte packets\\
       \hline Delay to send            & 0.23~ms, $\sigma$=0.046~ms & 0.23~ms, $\sigma$=0.021~ms \\
       \hline Delay to receive         &0.13~ms, $\sigma$=0.007~ms & 0.14~ms, $\sigma$=0.007~ms\\
       \hline
       \end{tabular}
   \end{center}
}
\end{table}

\begin{table}[t!]
{\small \topcaption{Average delays measured using Prawn on a mini-PC.}
\label{tab:minipc} \vspace{-5mm}
   \begin{center}
       \begin{tabular}{|l|c|c|}
       \hline \textbf{} & 100-byte packets  & 1,400-byte packets\\
       \hline Delay to send            &1.09~ms, $\sigma$=0.038~ms&1.15~ms, $\sigma$=0.043~ms\\
       \hline Delay to receive         &0.32~ms, $\sigma$=0.015~ms & 0.38~ms, $\sigma$=0.022~ms\\
       \hline
       \end{tabular}
   \end{center}
}
\end{table}

\begin{table}[t!]
{\small \topcaption{Average delays measured using Prawn on a Nokia
N770.} \label{tab:nokia} \vspace{-5mm}
   \begin{center}
       \begin{tabular}{|l|c|c|}
       \hline \textbf{} & 100-byte packets  & 1,400-byte packets\\
       \hline Delay to send            &3.09~ms, $\sigma$=0.21~ms&4.22~ms, $\sigma$=0.22~ms\\
       \hline Delay to receive         &1.47~ms $\sigma$=0.22~ms & 1.61~ms, $\sigma$=0.27~ms\\
       \hline
       \end{tabular}
   \end{center}
}
\end{table}

\vspace{3mm}

\section{Performance of Prawn}
\label{sec:overhead}

In this section, we present our measurements of the overhead
introduced by Prawn (in terms of delay and throughput) on the
different platforms that compose our testbed. We show that Prawn
delivers adequate performance even in the case of platforms with
limited computation-- and memory capability.

\subsection{Setup}

The experiments reported here were performed using different platforms
from our testbed, namely:

\begin{itemize}

\item \textbf{Laptop 1}. Dell Latitude X1 featuring and Intel
Pentium~M 733 Processor at 1.1~GHz, 1.2~GB of memory and an embedded
Intel PRO/Wireless 2200BG 802.11b/g chipset. The Operating System
(OS) is Linux Fedora Core 6, 2.6.18.2 kernel, with the ipw2200
driver.

\item \textbf{Laptop 2}. HP Compaq nx7000 featuring an Intel Pentium~M
Processor at 1.4~GHz, 512~MB of memory and a Netgear WG511T
802.11b/g wireless Cardbus adapter. The operating system is Linux
Fedora Core 6, 2.6.18.2 kernel, with the madwifi-ng driver for
Atheros chipsets.

\item \textbf{Mini-Pc}. VIA Eden EBGA fanless
processor at 600~MHz, with 512~MB of memory and a Cisco
Aironet 802.11a/b/g wireless PCI adapter (PI21AG-E-K9).
The OS is Linux Fedora Core
5, 2.6.16.16 kernel, with the madwifi-ng driver for Atheros
chipsets.

\item \textbf{PDA}. Nokia N770 Internet Tablet, powered by a 250~MHz
ARM based Texas Instruments 1710 OMAP processor, 64~MB of memory and
an embedded 802.11b/g chipset. The OS used is the
Nokia Internet Tablet 0S2006 Edition.
\end{itemize}

Our experiments were performed in a research laboratory (i.e.,
indoors, under radio interference of existing wireless networks,
etc.). Measurements were obtained between two nodes forming a one-hop
topology.

\subsection{Average Delay}

When sending a message, a prototype using Prawn calls the ${\tt
Prawn\_Send}()$ primitive. This function forwards the packet to the
engine, which triggers an event. If not idle, the engine terminates
its current tasks (e.g., sending/receiving a beacon, receiving a
packet) and encapsulates the message. Then it changes the transmit
power (if requested) and sends the packet to the corresponding
neighbor through the wireless interface. Our goal here is to measure
the additional delay incurred by Prawn (without Prawn, a message would
be sent directly to the wireless interface) and show that it is small
enough compared to the overall message delivery delay.

To measure the additional delay incurred by Prawn, we implemented a
simple application program that calls the ${\tt Prawn\_Send}()$
primitive and records a timestamp. Then the Prawn Engine receives the
packet from the loop-back interface, encapsulates it, changes the
transmit power if requested, and generates a second timestamp just
before the packet is finally sent through the wireless interface.
Experiments have been performed for two different packet sizes, namely
100-- and 1,400 bytes.

Tables~\ref{tab:laptop2} and~\ref{tab:laptop1} show the results when
running Prawn on our testbed laptops. Reported averages were obtained
over 10,000 measures. These results are quite encouraging as
additional delays of up to 0.20~ms for sending a message are quite
reasonable for the purposes of a prototype.

Measured delays for the the mini-PC and the Nokia N770 are presented
in Tables~\ref{tab:minipc} and~\ref{tab:nokia}. As expected, the
delays are higher, but still sufficiently non-intrusive when observing
the behavior of routing protocols or addressing mechanisms with real
wireless links and users.

\begin{table}[t!]
\topcaption{Average throughput using Prawn on a Dell Latitude X1
laptop.} {\small \label{tab:dell-throughput} \vspace{-5mm}
    \begin{center}
        \begin{tabular}{|l|c|c|}
        \hline  & 100-byte packets & 1,400-byte packets\\
        \hline \hline With Prawn            & 3.0 Mbit/s & 17.9 Mbits/s \\
        \hline Without Prawn           & 3.1 Mbit/s & 18 Mbit/s \\
        \hline
        \end{tabular}
    \end{center}
    }
\end{table}

\begin{table}[t!]
\topcaption{Average throughput using Prawn on a HP Compaq nx7000 laptop.} {\small
\label{tab:hp-throughput} \vspace{-5mm}
    \begin{center}
        \begin{tabular}{|l|c|c|}
        \hline  & 100-byte packets & 1,400-byte packets\\
        \hline \hline With Prawn            & 4.7 Mbit/s & 25.6 Mbits/s \\
        \hline Without Prawn           & 4.9 Mbit/s & 25.9 Mbit/s \\
        \hline
        \end{tabular}
    \end{center}
}
\end{table}

\subsection{Throughput}
\label{subsec:throughput}

We also measured the maximum throughput supported by Prawn.  To this
end, we compared the throughput of two Perl scripts communicating by
sending data directly through the wireless interface (i.e., without
Prawn), against having them send data using Prawn's send/receive
primitives.  Results for the laptop nodes are presented in
Tables~\ref{tab:dell-throughput}
and~\ref{tab:hp-throughput}.\footnote{Recall that IEEE~802.11a/g (we
used 802.11g in our experiments) has a maximum nominal transmission
rate of 54~Mbps. Actually, this is the rate for the payload of
802.11 frames; headers, trailers, and handshake packets are sent at
lower rates, which makes the effective throughput drop below
40~Mbps.}

These results are again very encouraging: the throughput achieved with
Prawn is comparable to what was obtained without Prawn, which shows
that the overhead introduced by Prawn is mostly negligible. However,
as shown in Tables~\ref{tab:minipc-throughput}
and~\ref{tab:nokia-throughput}, this is not the case for the less
capable platforms, namely the mini-PC and Nokia N770 which exhibit
already very low throughput without Prawn. The per-packet processing
performed by Prawn increases the bottleneck, limiting throughput further.

Nevertheless, we argue that when testing protocol functionality and
correctness, achieving high throughput is not critical. Indeed, most
target protocols such as routing, topology control, localization,
addressing/naming schemes, etc. generate typically short control
messages relatively sparse in time.

\begin{table}[t!]
\topcaption{Average throughput of mini-PC.} {\small
\label{tab:minipc-throughput} \vspace{-5mm}
    \begin{center}
        \begin{tabular}{|l|c|c|}
        \hline  & 100-byte packets & 1,400-byte packets\\
        \hline \hline With Prawn            & 2.2 Mbit/s & 11.6 Mbits/s \\
        \hline Without Prawn           & 5.4 Mbit/s & 22.2 Mbit/s \\
        \hline
        \end{tabular}
    \end{center}
}
\end{table}

\begin{table}[t!]
\topcaption{Average throughput of Nokia N770.} {\small
\label{tab:nokia-throughput} \vspace{-5mm}
    \begin{center}
        \begin{tabular}{|l|c|c|}
        \hline  & 100-byte packets & 1,400-byte packets\\
        \hline \hline With Prawn            & 0.3 Mbit/s & 2.3 Mbits/s \\
        \hline Without Prawn           & 0.8 Mbit/s & 4.9 Mbit/s \\
        \hline
        \end{tabular}
    \end{center}
}
\end{table}

\subsection{Communication Overhead}
\label{subsec:estimating}

The communication overhead incurred by Prawn is due to its beacons,
feedback messages, as well as additional message headers.  This
overhead can be easily estimated as shown in the following
example. Let us consider a Prawn prototype running on a 6-node
wireless network where nodes are all in range of one another. The
beacon period is set to the default value of $5,000$~ms, and $5$
different transmit power levels are used for the beacons.  Thus,
during a beacon period each node: (1) broadcasts $5$ beacons (one for
each transmit power) and (2) broadcasts $5$ beacon replies (one for
each neighbor).  We also have to account for an additional 4
(header) bytes per packet.

For the case that the prototyped protocol sends $10$ data packets per
second, during a beacon period, a node sends a total of $5$ beacons,
$5$ beacon replies, and $50$ data packets. The corresponding overhead
is $120$~bytes for the beacons, $80$~bytes for the beacon replies, and
$200$~bytes for data packet headers. We obtain $400$~bytes of overhead
per beacon period, or $640$ bits/s.  Since we have $6$~nodes in the
network, the overhead for the whole network is $3.84$~Kbit/s.

The important point here is that, as long as protocol designers
understand the cost incurred by Prawn, they can, besides testing their
prototype under real conditions, also conduct absolute/relative
overhead analysis.

\section{Prototyping with Prawn}
\label{sec:case}

Prawn is intended to be a tool for prototyping a wide range of
communication algorithms for heterogeneous wireless networks.  In this
section, we first illustrate the use of Prawn through a number of case
studies, highlighting its range of applicability and ease of use as
well as how it can be employed to evaluate and test protocols.

\subsection{Case Study 1: Neighborhood Monitoring}
\label{subsec:monitoring}

\begin{figure}[t!]
\centering
\includegraphics[scale=0.55]{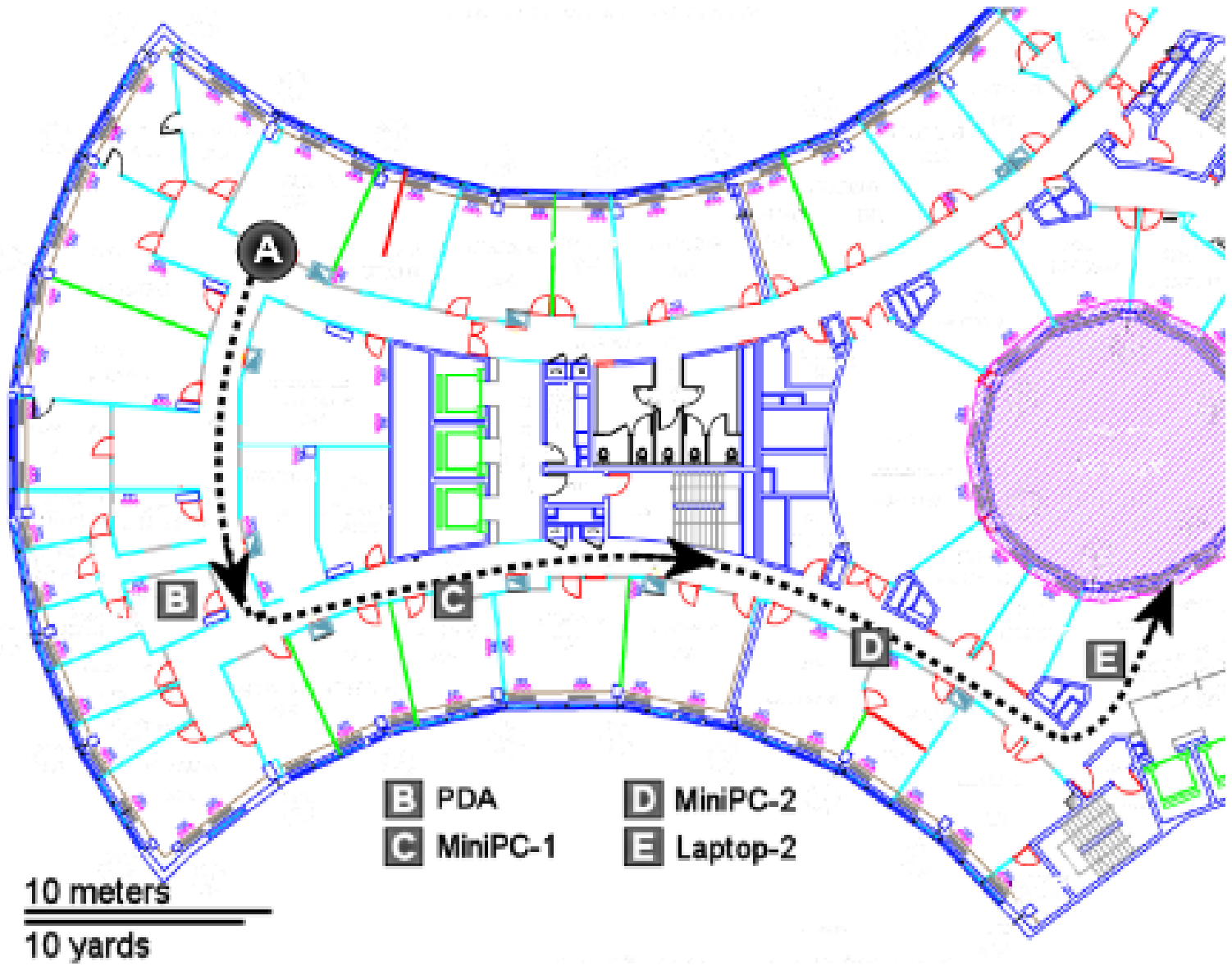}
\caption{Neighborhood monitoring experiment setup.}
\label{fig:labo}
\end{figure}

In this case study we use Prawn to implement a simple neighborhood
monitoring protocol. The purpose of this example is to show how Prawn
simplifies neighbor discovery and link quality assessment.  The
experimentation setup was as follows: four heterogeneous nodes running
Prawn were placed at different locations in our lab as depicted in
Figure~\ref{fig:labo}. This figure also lists the types of nodes (for
further details, see Section~\ref{sec:overhead}).

\begin{figure*}[t!]
\begin{center}
 \subfloat[Received signal strength measured from the mobile laptop while moving.]{
     \label{fig:laptop}
     \includegraphics[scale=0.60]{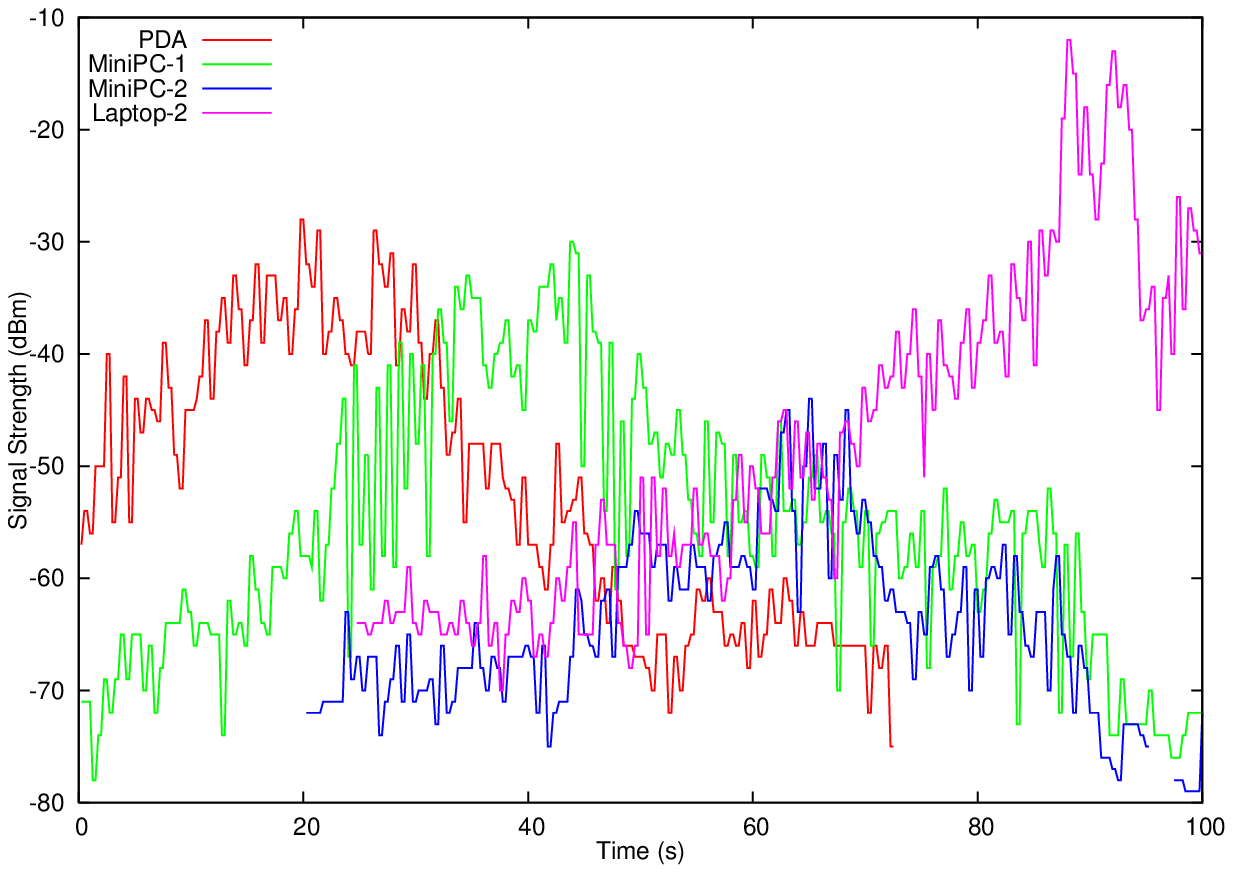}
 }~~~
\subfloat[NS-2 simulation of the received signal strength from the
mobile laptop.]{
    \label{fig:simu}
    \includegraphics[scale=0.60]{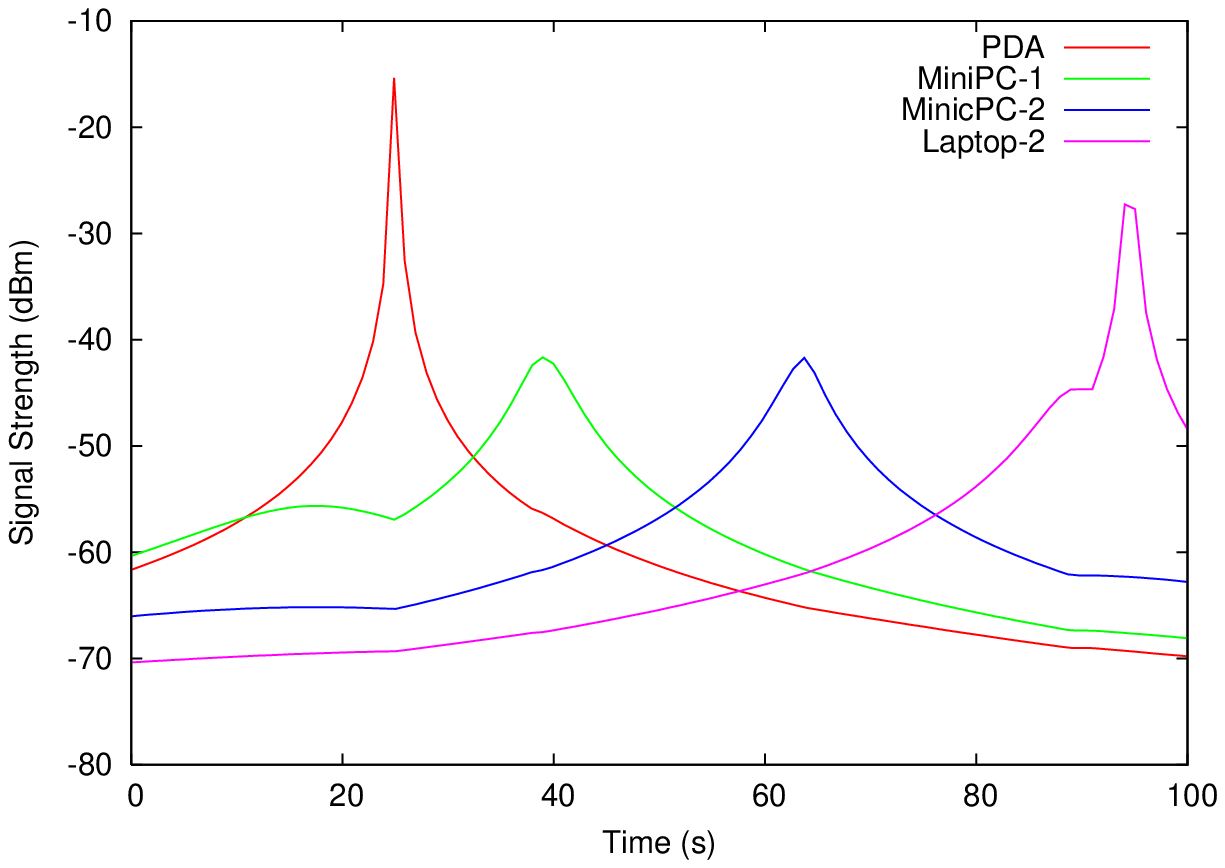}
} \caption{Received signal strength for the experiment with a mobile
node and four static nodes.}\label{fig:laptop-simu}
\end{center}
\end{figure*}

We used a fifth node running Prawn executing a simple script
registering the received signal strength from each static node. This
fifth node was a laptop that we moved along the corridor at walking
speed. Figure~\ref{fig:laptop} plots the received signal strength
data collected by the moving node. It is interesting to remark how
at the beginning of the experiment the laptop has only two direct
neighbors: the PDA and the MiniPC-1. Indeed the curves for the other
two nodes appear only after 20 seconds. The same information can be
deduced from Figure~\ref{fig:list-engine-b}, which depicts a
snapshot of the feedback received by Prawn in console mode. The same
figure shows that the other two nodes (MiniPC-2 and Laptop-2) are in
the two-hop neighborhood of the moving laptop. It can be also
observed that at the end of the experiment the laptop has only three
direct neighbors; indeed the PDA becomes too far and its curve in
Figure~\ref{fig:laptop} ends after around 75 seconds.  This is also
confirmed in Figure~\ref{fig:list-engine-e}, which lists all the
fixed nodes but has the PDA marked as dead.\footnote{The reader is
referred back to Section~\ref{sec:engine} for more details on how
Prawn assumes that a node is no longer available.}

\begin{figure}[t!]
  \begin{center}
{\tiny
\begin{verbatim}
       ================ Neighbor List for node Laptop-1 ================
       =================================================================
       89C2  Active  00:14:A7:FA:89:C2      Beacon period : 1000  PDA
             Weakest beacon received by this neighbor : 100 mW
             -----------------------------------------------------------
       100mW  Active  @1020  R 0  S 5  B 5 [3.9810717 nW (-54 dBm)]  5/5

       2hop-Neighbors:  MiniPC-2 (100 mW, - 78 dBm)
                        MiniPC-1 (100 mW, - 61 dBm)
       =================================================================

       =================================================================
       3051  Active  00:40:96:A7:30:51    Beacon period : 1000  MiniPC-1
             Weakest beacon received by this neighbor : 100 mW
             -----------------------------------------------------------
       100mW  Active  @1020  R 0  S 9  B 9 [0.0158489 nW (-78 dBm)]  4/5

       2hop-Neighbors:  PDA (100 mW, -50 dBm) MiniPC-2 (100 mW, -56 dBm)
       =================================================================
\end{verbatim}
} \caption{Snapshot of Prawn running in console mode on the mobile
laptop at the beginning of the experiment.} \label{fig:list-engine-b}
\end{center}
\end{figure}

\begin{figure}[t!]
  \begin{center}
{\tiny
\begin{verbatim}
       ================ Neighbor List for node Laptop-1 ================
       =================================================================
       89C2  Dead     00:14:A7:FA:89:C2     Beacon period : 1000  PDA
             Weakest beacon received by this neighbor : 100 mW
             -----------------------------------------------------------
       100mW  Dead  @1082  R 0  S 73  B 70 [0.0158489 nW (-78 dBm)]  0/5

       2hop-Neighbors:  MiniPC-2 (100 mW, - 75 dBm)
                        MiniPC-1 (100 mW, - 58 dBm)
       =================================================================

       =================================================================
       3051  Active   00:40:96:A7:30:51   Beacon period : 1000  MiniPC-1
             Weakest beacon received by this neighbor : 100 mW
             -----------------------------------------------------------
       100mW  Active  @1119  R 0  S 109  B 119 [0.0794 nW (-71 dBm)] 4/5

       2hop-Neighbors:  PDA (100 mW, -50 dBm) MiniPC-2 (100 mW, -55 dBm)
       =================================================================

       =================================================================
       2F6D  Active  00:40:96:A9:2F:6D   Beacon period : 1000   MiniPC-2
             Weakest beacon received by this neighbor : 100 mW
             -----------------------------------------------------------
       100mW  Active  @1119  R 1  S 134  B 139 [0.0794 nW (-71 dBm)] 3/5

       2hop-Neighbors:  PDA (lost)  MiniPC-1 (100 mW, -51 dBm)
                        Laptop-2 (100 mW, -55 dBm)
       =================================================================

       =================================================================
       3051  Active   00:40:96:A7:30:51   Beacon period : 1000  Laptop-2
             Weakest beacon received by this neighbor : 100 mW
             -----------------------------------------------------------
       100mW  Active @1120  R 0  S 197  B 86 [199.526 nW (-37 dBm)]  5/5

       2hop-Neighbors:  PDA (lost)  MiniPC-2 (100 mW, -54 dBm)
                        MiniPC-1 (100 mW, -75 dBm)
       =================================================================
\end{verbatim}
} \caption{Snapshot of Prawn running in console mode on the mobile
laptop at the end of the experiment.} \label{fig:list-engine-e}
\end{center}
\end{figure}

We compared the received signal strength results obtained with the
Prawn prototype against what is reported by simulations of the same
setup using NS-2~\cite{Bib:NS-2}. The discrepancy between the two sets
of results clearly illustrates the importance of testing wireless
protocols under real conditions. Prawn's value-added is that it
makes prototyping protocols as simple and fast as implementing them
on a network simulator.

\subsection{Case Study 2: Node Localization in Wireless Mesh Networks}
\label{subsec:localization}

\begin{figure}[!t]
\begin{center}
\includegraphics[width=\linewidth]{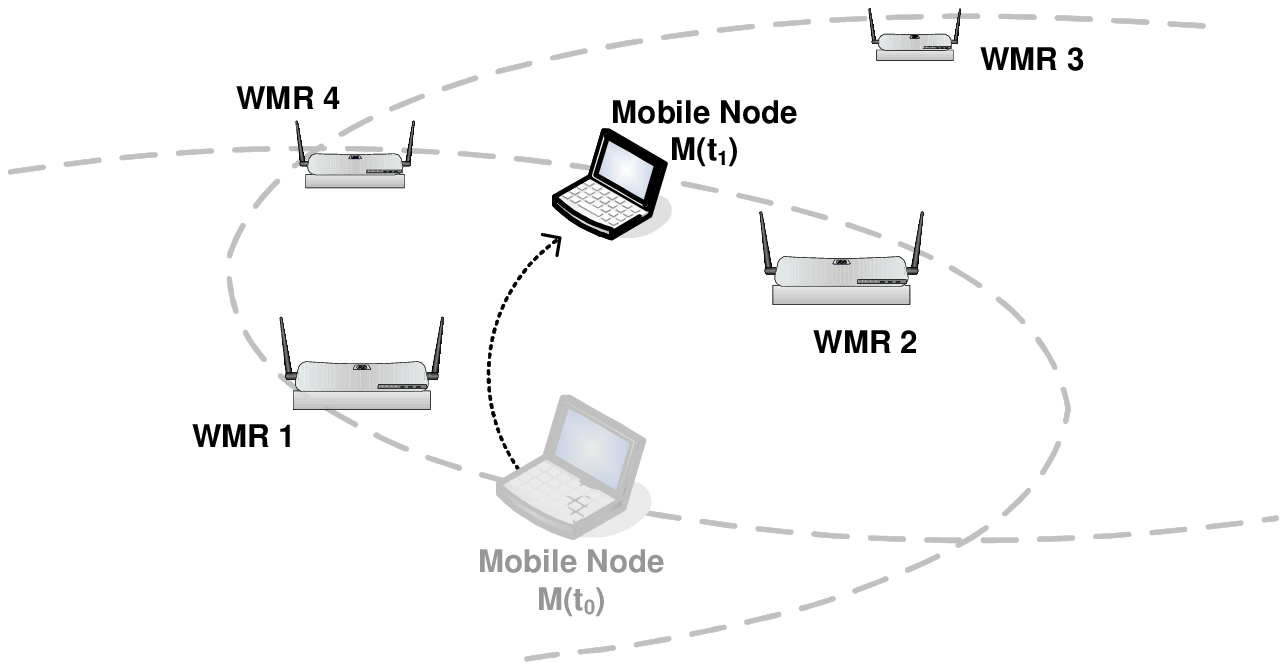}
\caption{Keeping track of moving nodes} \label{fig:feedbackB}
\end{center}
\end{figure}

The previous case study can be extended for Wireless Mesh Networks
(WMNs). In WMNs, mobile users connect to fixed wireless nodes
(Wireless Mesh Routers~-- WMRs) belonging to the infrastructure. As
illustrated in Figure~\ref{fig:feedbackB}, consider the case where
${\tt WMR1}$, ${\tt WMR2}$, and mobile node ${\tt M}$ are all
neighbors (i.e., they are in range of one another). Then node ${\tt
WMR1}$, can approximately determine the direction of movement of ${\tt
M}$ by measuring the received signal strength or the minimum transmit
power of beacons received. In this particular example, suppose that
${\tt WMR1}$ is experimenting decreasing link quality with ${\tt M}$.
${\tt WMR1}$ can use the feedback packets broadcast by another fixed
node, say ${\tt WMR2}$, to ${\tt M}$. By examining these packets,
${\tt WMR1}$ realizes that the link quality of ${\tt WMR2-M}$ is not
getting worse. Thus, ${\tt WMR1}$ concludes ${\tt M}$ is moving
approximately in the direction of ${\tt WMR2}$.  With this
information, a routing process can choose to start forwarding packets
to ${\tt WMR2}$ in order to reach ${\tt M}$. Note that this would also
be very useful in the context of episodically-connected
networks. Figure~\ref{fig:wmn} shows a simple script that uses
information from the Prawn Engine to find all neighbors of ${\tt M}$
sorting them by RSSI.

\begin{figure}[!t]
\begin{center}
 \input rssi_m.pl.tex\caption{Prototype of a simple node localization algorithm.} \label{fig:wmn}
\end{center}
\end{figure}

Furthermore, by using the maximum received signal strength (already
included in the feedback packets), a node can estimate its
neighborhood's ``virtual topology'', where distances between nodes are
based on signal strength (i.e., are given in mW or dBm). Note that
these ``virtual distances'' between nodes are not always proportional
to euclidian distances, e.g., if the nodes are not in
line-of-sight. Nevertheless, signal strength provides indeed a better
characterization of link quality than the physical distance between
nodes. By using simple trilateration (with 3 or more neighbors), nodes
can compute their locations more accurately. For the example
illustrated in Figure~\ref{fig:feedbackB}, node ${\tt WMR1}$ can
compute the location of ${\tt M}$ by trilateration, using its own
RSSI (Received Signal Strength Indication) on the link ${\tt M-WMR1}$,
along with the RSSIs on the links ${\tt M-WMR4}$ and ${\tt M-WMR2}$
(broadcast respectively by neighbor nodes ${\tt WMR4}$ and ${\tt
WMR2}$).

This particular example illustrates the use of Prawn's features
related to power control and signal measurements. It also show cases
the modularity properties of Prawn's prototypes, which enables code
reused. In other words, Prawn code to implement basic functions like
neighborhood discovery can be reused by other (more complex)
prototypes.

\subsection{Case Study 3: Prototyping Other Protocols}

\subsubsection{Flooding}
\label{subsubsec:flooding}

Flooding is the simplest possible routing algorithm. Its basic
operation is as follows: upon receiving a packet, each node sends it
once to all its neighbors.\footnote{Of course, more elaborated
variations of flooding exist, but here we consider it in its
simplest form.} Thus the only requirement to implement this
algorithm is to be able to receive and broadcast packets.

Prawn makes this algorithm easier to implement even for inexperienced
programmers as they are abstracted away from lower-level functions
like sockets, ports, addressing, etc. Flooding can be implemented
simply by using the ${\tt Prawn\_Receive}()$ and ${\tt
Prawn\_Send\_Broadcast}()$ functions.

Figure~\ref{fig:flooding} shows how short and simple the flooding
prototype using the Prawn Library is. This 12-line piece of code has
been running successfully on our testbed and we have conducted
intensive experimentation that has enabled us to understand the
behavior of the flooding algorithm under realistic conditions and
with real users. This behavior is not obvious and known a priori
from simulations; for example, Cavin \emph{et al.}  tried to
simulate the flooding~\cite{Bib:Cavin02} algorithm using three
different simulators namely, NS-2, OPNET, and GloMoSim, with exactly
the same parameters and scenarios. Surprisingly, the results were
considerably different, depending on the simulator used.

\begin{figure}[!t]
\begin{center}
\input floodingLIGHT.pl.tex \caption{Perl code of a flooding
prototype} \label{fig:flooding}
\end{center}
\end{figure}


\subsubsection{Network Coding}

While the previous section illustrates the use of Prawn to prototype
one of the simplest protocols, we show, in this section, that Prawn
can also be used to prototype more complex protocols. In particular,
we show case the use of Prawn to prototype network coding
algorithms~\cite{Bib:Ahlswede00,Bib:Katti06}. For example, COPE,
whose principles and implementation are described in
~\cite{Bib:Katti06}, is clearly rather complex. Our goal here is to
show that some evaluation of network coding proposals could be
easily done without requiring a fully-functional implementation of
the algorithm.

For clarity, we briefly explain the essence of network coding through
a very simple example. In traditional forwarding, when a node ${\tt
A}$ and a node ${\tt B}$ want to exchange data via a third node ${\tt
C}$, both send their packets to ${\tt C}$, and then ${\tt C}$ forwards
the packets to ${\tt A}$ and to ${\tt B}$.  Exchanging a pair of
packets requires 4 transmissions. Using network coding, instead of
sending separate packets to ${\tt A}$ and ${\tt B}$, node ${\tt C}$
combines (e.g., using the XOR function) both packets received from
${\tt A}$ and ${\tt B}$, and broadcasts the encoded packet. Since
${\tt A}$ knows the packet it has sent, it can decode the packet sent
by ${\tt B}$ (e.g. applying again the XOR function) from the encoded
packet received from ${\tt C}$. Similarly, ${\tt B}$ can decode the
packet sent by ${\tt A}$ from the same packet received from ${\tt
C}$. Thus, with this method, only 3 transmissions, instead of 4, are
required.

Using Prawn, we implemented a prototype of the algorithm described above.
As shown in the Perl code running on node ${\tt C}$
(Figure~\ref{fig:nc}), the first received packet is stored in a
standby variable (${\tt \$Stdby}$), then the next packet is stored
as ${\tt \$Msg}$. If the two stored packets are not received from
the same node, then they are XORed and broadcast. If, instead, both
packets are from the same node, it does not make sense to XOR them.
In this case, the packet stored in standby is sent as a normal
unicast packet, and the latest packet goes to the standby queue.

\begin{figure}[!t]
\begin{center}
\input ncLIGHT.pl.tex \caption{Perl code of a network coding
algorithm} \label{fig:nc}
\end{center}
\end{figure}

We also implemented a prototype of a traditional forwarding
algorithm. We compare both implementations to measure the
performance gains achieved by network coding when ${\tt A}$ sends
10,000 packets of 1,400 bytes to ${\tt B}$ and vice-versa. Without
network coding, the total amount of data transmitted was 54~MB on
both links. With network coding, only 44~MB were sent. With this
code as a starting point, network coding protocol designers can test
and tune their algorithms on real platforms under real conditions.

\subsubsection{Topology Control}
\label{subsec:topology}

Topology control algorithms require updated information about
neighbors. Selecting good neighbors is often beneficial for the
whole network. Prawn supports varied neighbors selection criteria
relying on cross-layer information. For instance, in order to save
energy and reduce interference, neighbors with lowest required
transmit power can be selected. Conversely, neighbors with the
highest signal strength received could be chosen. Many recent
research efforts relying on cross-layer approaches would benefit
from Prawn's lower layer information.

\begin{figure}[!t]
\begin{center}
\input topologyLIGHT.pl.tex \caption{Perl code of a topology
control prototype} \label{fig:topology}
\end{center}
\end{figure}

The code in Figure~\ref{fig:topology} shows how to get in 7 lines a
list of neighbors sorted according to their receive signal strength.
This code is running successfully on our testbed consisting of
heterogeneous nodes. An important point here is that the received
signal strength value retrieved from the wireless driver can be
different depending on the wireless device model. If the neighbors
do not have all the same wireless cards, the selection could be
biased. This is an example of practical issue that cannot be taken
into account from simulations. Using Prawn, designers can evaluate
their proposal taking into account the features and performance of
off-the-shelf hardware and drivers.

\section{Conclusion}
\label{sec:conclusion}

In this paper we proposed Prawn, a novel prototyping tool for
high-level network protocols and applications. Prawn's main goal is
to facilitate the prototyping of wireless protocols so that
prototyping becomes an integral part of the design process of
wireless systems.

Prawn is not an alternative to simulation or any other evaluation
method. Instead, it stands as a complementary approach that goes
beyond simulation by taking into account real-world properties.
Prawn surfs the wave of recent research efforts toward making
implementation easier (e.g., Click and XORP), but as a preliminary
phase in this process. The designer has to keep in mind, however,
that the performance of a prototype does not always match exactly
with the performance of a final and optimized implementation.
Nevertheless, it is not the same gap we can observe between
simulation results and real implementation results. In simulation it
is very difficult to estimate how far a model is from reality and
the exact impact it has on the performance results. Using Prawn, an
estimation (even rough) can be deduced from the observed overhead.

Unlike existing implementation tools, Prawn provides a general,
simple, concise, yet sufficient set of functions for a wide range of
high-level algorithms, as well as an API that shields the designer
from low-level implementation details. Through several case studies,
we showcased the use of Prawn in the context of a wide range of
network protocols. But the possibilities of Prawn are not restricted
to the examples given in this paper. Other experiments where Prawn
can be useful include: evaluating existing protocols for wired
networks in the wireless context, implementing new routing
protocols, testing overlay approaches in wireless multi-hop
networks, evaluating distributed security algorithms, testing new
naming mechanisms over IP, testing incentive mechanisms for
communities, implementing localization algorithms, measuring
wireless connectivity in both indoor and outdoor scenarios,
evaluating peer-to-peer algorithms, testing opportunistic forwarding
mechanisms.

We hope our work will provide a starting point for an improved
design methodology as prototyping provides both easy and accurate
evaluation of wireless protocols and services under real conditions.
This paper has demonstrated that this is feasible~-- Prawn is a
fully-functional tool that responds to the needs of early protocol
evaluation. Finally, we expect that Prawn's simplicity will allow
researchers to adopt it. To help this becoming true, ongoing work
includes adding new prototyping facilities, releasing a Java version
of the Prawn Library, and porting Prawn to other operating systems
such as FreeBSD and Windows.

\bibliographystyle{IEEEtran}
\bibliography{ton07}

\begin{thebibliography}{10}
\providecommand{\url}[1]{#1}
\csname url@rmstyle\endcsname
\providecommand{\newblock}{\relax}
\providecommand{\bibinfo}[2]{#2}
\providecommand\BIBentrySTDinterwordspacing{\spaceskip=0pt\relax}
\providecommand\BIBentryALTinterwordstretchfactor{4}
\providecommand\BIBentryALTinterwordspacing{\spaceskip=\fontdimen2\font plus
\BIBentryALTinterwordstretchfactor\fontdimen3\font minus
  \fontdimen4\font\relax}
\providecommand\BIBforeignlanguage[2]{{%
\expandafter\ifx\csname l@#1\endcsname\relax
\typeout{** WARNING: IEEEtran.bst: No hyphenation pattern has been}%
\typeout{** loaded for the language `#1'. Using the pattern for}%
\typeout{** the default language instead.}%
\else
\language=\csname l@#1\endcsname
\fi
#2}}

\bibitem{Bib:Raychaudhuri05}
D.~Raychaudhuri, M.~Ott, and I.~Seskar, ``Orbit radio grid tested for
  evaluation of next-generation wireless network protocols.'' in
  \emph{Proceedings of Tridentcom}, 2005, pp. 308--309.

\bibitem{Bib:Zhu05}
W.~Zhu, D.~Browne, and M.~P. Fitz, ``An open access wideband multi-antenna
  wireless testbed with remote control capability.'' in \emph{Proceedings of
  Tridentcom}, 2005, pp. 72--81.

\bibitem{Bib:White03}
B.~White, J.~Lepreau, and S.~Guruprasad, ``Lowering the barrier to wireless and
  mobile experimentation,'' \emph{Computer Communications Review}, vol.~33,
  no.~1, pp. 47--52, Jan. 2003.

\bibitem{Bib:De06}
P.~De, A.~Raniwala, R.~Krishnan, K.~Tatavarthi, J.~Modi, N.~A. Syed, S.~Sharma,
  and T.~cker Chiueh, ``Mint-m: an autonomous mobile wireless experimentation
  platform.'' in \emph{Proceedings of {ACM/USENIX Mobisys}}, 2006, pp.
  124--137.

\bibitem{Bib:roofnet}
\BIBentryALTinterwordspacing
{MIT Computer Science and Artificial Intelligence Laboratory (CSAIL)}. {MIT}
  roofnet. [Online]. Available:
  \url{http://pdos.csail.mit.edu/roofnet/doku.php}
\BIBentrySTDinterwordspacing

\bibitem{Bib:Kohler00}
E.~Kohler, R.~Morris, B.~Chen, J.~Jannotti, and M.~F. Kaashoek, ``The {C}lick
  modular router,'' \emph{ACM Trans. Comput. Syst.}, vol.~18, no.~3, pp.
  263--297, 2000.

\bibitem{Bib:Handley03}
M.~Handley, O.~Hodson, and E.~Kohler, ``{XORP}: an open platform for network
  research,'' \emph{Computer Communications Review}, vol.~33, no.~1, pp.
  53--57, 2003.

\bibitem{Bib:NS-2}
\BIBentryALTinterwordspacing
{The Network Simulator NS-2}. [Online]. Available:
  \url{http://www.isi.edu/nsnam/ns/}
\BIBentrySTDinterwordspacing

\bibitem{Bib:OPNET}
\BIBentryALTinterwordspacing
{OPNET Modeler}. [Online]. Available:
  \url{http://www.opnet.com/products/modeler/}
\BIBentrySTDinterwordspacing

\bibitem{Bib:Glomosim}
X.~Zeng, R.~Bagrodia, and M.~Gerla, ``{GloMoSim}: a library for parallel
  simulation of large-scale wireless networks,'' in \emph{Proceedings of
  Workshop on Parallel and Distributed Simulation}, Banff, Canada, May 1998.

\bibitem{Bib:Qualnet}
\BIBentryALTinterwordspacing
{QualNet Simulator}. [Online]. Available: \url{http://www.qualnet.com}
\BIBentrySTDinterwordspacing

\bibitem{Bib:Zheng03}
P.~Zheng and L.~Ni, ``{EMPOWER:} {A} network emulator for wireless and wireline
  networks,'' in \emph{Proceedings of {IEEE Infocom}}, San Francisco, CA, Apr.
  2003.

\bibitem{Bib:Kojo01}
M.~Kojo, A.~Gurtov, J.~Manner, P.~Sarolahti, T.~O. Alanko, and K.~E.~E.
  Raatikainen, ``Seawind: a wireless network emulator,'' in \emph{GI/ITG
  Conference on Measuring, Modelling and Evaluation of Computer and
  Communication Systems}, Aachen, Germany, Sept. 2001.

\bibitem{Bib:Ramachandran05}
K.~Ramachandran, S.~Kaul, S.~Mathur, M.~Gruteser, and I.~Seskar, ``Towards
  large-scale mobile network emulation through spatial switching on a wireless
  grid,'' in \emph{Proceedings of {ACM Sigcomm}}, Philadelphia, PA, Aug. 2005.

\bibitem{Bib:bavier.sigcomm06}
A.~Bavier, N.~Feamster, M.~Huang, L.~Peterson, and J.~Rexford, ``In {VINI
  Veritas}: Realistic and controlled network experimentation,'' in
  \emph{Proceedings of {ACM Sigcomm}}, Pisa, Italy, Sept. 2006.

\bibitem{Bib:PlanetLab}
\BIBentryALTinterwordspacing
Planetlab: An open platform for developing, deploying, and accessing
  planetary-scale services. [Online]. Available:
  \url{http://www.planet-lab.org}
\BIBentrySTDinterwordspacing

\bibitem{Bib:touch.global98}
J.~Touch and S.~Hotz, ``The {X-Bone},'' in \emph{Global Internet
  Mini-Conference}, Sidney, Australia, Nov. 1998.

\bibitem{Bib:jiang.ispdpa04}
X.~Jiang and D.~Xu, ``Violin: Virtual internetworking on overlay
  infrastructure,'' in \emph{Proceedings of the International Symposium on
  Parallel and Distributed Processing and Applications}, Hong-Kong, China, Dec.
  2004.

\bibitem{Bib:eide.nsdi07}
E.~Eide, L.~Stoller, and J.~Lepreau, ``An experimentation workbench for
  replayable networking research,'' in \emph{Usenix Symposium on Networked
  Systems Design and Implementation}, Cambridge, MA, Apr. 2007.

\bibitem{Bib:Polastre05}
J.~Polastre, J.~Hui, P.~Levis, J.~Zhao, D.~E. Culler, S.~Shenker, and
  I.~Stoica, ``A unifying link abstraction for wireless sensor networks.'' in
  \emph{Proceedings of {ACM Sensys}}, 2005, pp. 76--89.

\bibitem{Bib:Tinyos}
P.~Levis, S.~Madden, J.~Polastre, R.~Szewczyk, K.~Whitehouse, A.~Woo, D.~Gay,
  J.~Hill, M.~Welsh, E.~Brewer, and D.~Culler, ``Tiny{OS}: An operating system
  for wireless sensor networks,'' in \emph{Ambient Intelligence}.\hskip 1em
  plus 0.5em minus 0.4em\relax Springer-Verlag, 2004.

\bibitem{Bib:Girod04}
L.~Girod, J.~Elson, A.~Cerpa, T.~Stathopoulos, N.~Ramanathan, and D.~Estrin,
  ``Emstar: A software environment for developing and deploying wireless sensor
  networks.'' in \emph{Proceedings of {USENIX Annual Technical Conference}},
  2004, pp. 283--296.

\bibitem{Bib:XBow2}
\BIBentryALTinterwordspacing
{Crossbow}. Intel stargate. [Online]. Available: \url{http://www.xbow.com}
\BIBentrySTDinterwordspacing

\bibitem{Bib:Mapi}
\BIBentryALTinterwordspacing
M.~Youssef. {MAPI: An API for wireless cards under linux}. [Online]. Available:
  \url{http://www.cs.umd.edu/˜moustafa/}
\BIBentrySTDinterwordspacing

\bibitem{Bib:Jt}
\BIBentryALTinterwordspacing
{Wireless Tools for Linux}. [Online]. Available:
  \url{http://hpl.hp.com/personal/Jean\_Tourrilhes/ Linux/Tools.html}
\BIBentrySTDinterwordspacing

\bibitem{Bib:prawn}
\BIBentryALTinterwordspacing
Prawn website. [Online]. Available: \url{http://prawn.lip6.fr/}
\BIBentrySTDinterwordspacing

\bibitem{Bib:blough.tmc06}
D.~M. Blough, M.~Leoncini, G.~Resta, and P.~Santi, ``The k-neighbors approach
  to interference bounded and symmetric topology control in ad hoc networks,''
  \emph{IEEE Transactions on Mobile Computing}, vol.~5, pp. 1267--1282, Sept.
  2006.

\bibitem{Bib:li.ton05}
N.~Li and J.~Hou, ``Localized topology control algorithms for heterogeneous
  wireless networks,'' \emph{IEEE/ACM Transactions on Networking}, vol.~13, pp.
  1313--1324, Dec. 2005.

\bibitem{Bib:Cavin02}
D.~Cavin, Y.~Sasson, and A.~Schiper, ``On the accuracy of manet simulators,''
  in \emph{In proceedings of POMC'02}, Toulouse, France, Oct. 2002.

\bibitem{Bib:Ahlswede00}
R.~Ahlswede, N.~Cai, S.-Y.~R. Li, and R.~W. Yeung, ``Network information
  flow,'' \emph{IEEE Transactions on Information Theory}, vol.~46, no.~4, pp.
  1204--1216, July 2000.

\bibitem{Bib:Katti06}
S.~Katti, H.~Rahul, W.~Hu, D.~Katabi, M.~M{\'e}dard, and J.~Crowcroft, ``{XORs}
  in the air: practical wireless network coding,'' in \emph{Proceedings of {ACM
  Sigcomm}}, Pisa, Italy, 2006.

\end{thebibliography}

\end{document}